\documentclass[conference]{IEEEtran}
\IEEEoverridecommandlockouts
\usepackage{}
\usepackage{amsmath,amssymb,amsfonts}
\usepackage{algorithmic}
\usepackage{graphicx}
\usepackage{textcomp}
\usepackage[table]{xcolor}
\usepackage{multirow}
\usepackage{array}
\usepackage{url} 
\usepackage{verbatim}
\usepackage{subfig}

\def\BibTeX{{\rm B\kern-.05em{\sc i\kern-.025em b}\kern-.08em
    T\kern-.1667em\lower.7ex\hbox{E}\kern-.125emX}}
    
\usepackage[normalem]{ulem}
\usepackage{caption}
\DeclareCaptionFont{xipt}{\fontsize{9}{10}\mdseries}
\usepackage[font=xipt,labelfont=bf]{caption}

\definecolor{ao(english)}{rgb}{0.0, 0.5, 0.0}
\definecolor{powderblue(web)}{rgb}{0.69, 0.88, 0.9}
\definecolor{timberwolf}{rgb}{0.86, 0.84, 0.82}
\definecolor{wheat}{rgb}{0.96, 0.87, 0.7}
\definecolor{paleaqua}{rgb}{0.74, 0.83, 0.9}
\definecolor{silver}{rgb}{0.75, 0.75, 0.75}
\newcommand{\gh}[1]{\textcolor{blue}{#1}}

\newcolumntype{P}[1]{>{\raggedright \arraybackslash}p{#1}}

\begin{document}
\urlstyle{tt}
\title{Multi-channel U-Net for Music Source Separation
\thanks{
This work has received funding from the European Union's Horizon 2020 research and innovation programme under the Marie Sk\l{}odowska-Curie grant agreement No. 713673.
V. S. K. has received financial support through “la Caixa” Foundation (ID 100010434), fellowship code: LCF/BQ/DI18/11660064. 
Additional funding comes from the MICINN/FEDER UE project with reference PGC2018-098625-B-I00, H2020-MSCA-RISE-2017 project with reference 777826 NoMADS, 
Spanish Ministry of Economy and Competitiveness under the María de Maeztu Units of Excellence Program (MDM-2015-0502) and the Social European Funds. We also thank Nvidia for the donation of GPUs.}
}


\author{
    \IEEEauthorblockN{Venkatesh S. Kadandale\IEEEauthorrefmark{1}, Juan F. Montesinos\IEEEauthorrefmark{1}, Gloria Haro\IEEEauthorrefmark{1}, Emilia G\'{o}mez\IEEEauthorrefmark{1}\IEEEauthorrefmark{2}}
    \IEEEauthorblockA{\IEEEauthorrefmark{1} Department of Information and Communications Technologies, Universitat Pompeu Fabra, Barcelona, Spain}
    \IEEEauthorblockA{\IEEEauthorrefmark{2} Joint Research Centre, European Commission, Seville, Spain
    \\\{venkatesh.kadandale, juanfelipe.montesinos, gloria.haro, emilia.gomez\}@upf.edu}
}

\maketitle

\begin{abstract}
A fairly straightforward approach for music source separation is to train independent models, wherein each model is dedicated for estimating only a specific source. Training a single model to estimate multiple sources generally does not perform as well as the independent dedicated models. However, Conditioned U-Net (C-U-Net) uses a control mechanism to train a single model for multi-source separation and attempts to achieve a performance comparable to that of the dedicated models. We propose a multi-channel U-Net (M-U-Net) trained using a weighted multi-task loss as an alternative to the C-U-Net. We investigate two weighting strategies for our multi-task loss: 1) Dynamic Weighted Average (DWA), and 2) Energy Based Weighting (EBW). DWA determines the weights by tracking the rate of change of loss of each task during training. EBW aims to neutralize the effect of the training bias arising from the difference in energy levels of each of the sources in a mixture. Our methods provide three-fold advantages compared to C-U-Net: 1) Fewer effective training iterations per epoch, 2) Fewer trainable network parameters (no control parameters), and 3) Faster processing at inference. Our methods achieve performance comparable to that of C-U-Net and the dedicated U-Nets at a much lower training cost.

\end{abstract}


\section{Introduction}
Music source separation is the automatic estimation of the individual isolated sources that make up the audio mixture. It has been one of the most popular research problems in the music information retrieval community. Since most of the music audio present in the world exists in the form of mixtures, there are several applications of a system capable of music source separation -- e.g. automatic creation of karaoke, music transcription, music unmixing and remixing, music production and assistance in music education.

\begin{figure}[ht]%
 \centering
 \subfloat[Dedicated Models]{\includegraphics[width=55mm]{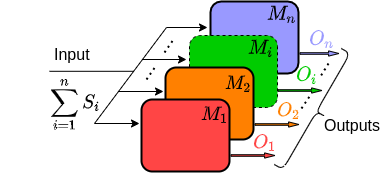} \label{fig:a}}%
 \hspace*{-8mm}
 \subfloat[Conditioned Model]{\includegraphics[scale=0.41]{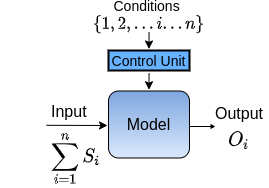}\label{fig:b}}
 \\
 \subfloat[Multi-task Model]{\includegraphics[scale=0.5]{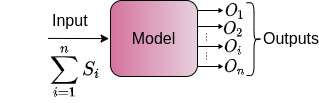}\label{fig:c}}
\caption{Typical models for music source separation.}
\label{fig1}
\end{figure}

We are interested in training a system discriminatively to estimate the sources present in the audio mixture. The deep neural networks (DNN) have been extensively used for this purpose. The existing methods mostly use DNN with either the spectrogram as the input signal representation \cite{nugraha2016multichannel,grais2016combining, roma2018improving} or directly the time-domain representation \cite{stoller2018wave,defossez2019demucs} to train such a system. The spectrograms are compact representations of time-domain waveforms. The networks operating directly on the time-domain waveforms require larger convolution kernels than those operating on the spectrograms because of the higher time resolution in the time-domain waveforms. Hence, the number of trainable network parameters in the waveform based models are generally higher than that of spectrogram based models. In this way, the spectrogram based models have lesser training cost than the waveform based ones. In most of the spectrogram based methods, the networks are trained to estimate masks, like binary or ratio masks. These masks are then multiplied with the magnitude spectrogram of the mixture to obtain the estimates of the corresponding sources. 

Convolutional neural networks (CNN) \cite{grais2016combining,park2018music} and long short term memory (LSTM) \cite{seetharaman2019class,stoter2019open} networks are the popular choices for DNN model architectures adapted for music source separation. Some of the latest top performing music source separation models are Open-Unmix \cite{stoter2019open}, MMDenseLSTM \cite{takahashi2018mmdenselstm}, Demucs \cite{defossez2019demucs} and Meta TasNet \cite{samuel2020meta}. While Open-Unmix and MMDenseLSTM models comprise of LSTMs and operate on spectrogram input, the other two methods operate directly on the time-domain waveform. It is not possible to single out any one of these models as the best model because they differ in number of trainable parameters, training time and performance metrics with respect to each of the sources in the mixture. Li et al. propose Sams-Net \cite{Li2019SamsNetAS} for music source separation which uses attention mechanism along with CNN layers and achieves larger receptive field than CNNs and LSTMs. Considering the low number of trainable parameters in the network, Sams-Net performs remarkably better than most of the other music source separation methods. Among the CNN based methods operating on spectrogram input, the U-Net \cite{ronneberger2015u} based methods \cite{jansson2017singing,stoller2018adversarial,pretet2019singing} have been popular owing to their simplicity and ease of training.

Music source separation systems can be categorized into one of the three variants shown in Fig.~\ref{fig1}. Type (a) system consists of independently trained models which do not share any trainable network parameters with each other. Each model is dedicated for estimating only a particular source. In such systems, the number of trainable network parameters increases proportionally to the number of sources to be separated. On the other hand, a type (b) system contains a single model regardless of the number of sources to be estimated. See \cite{meseguer2019conditioned} for example. In this case, the model outputs the estimate of a particular source based on the applied conditioning. As a consequence, for training type (b) systems, a data sample needs to be passed through the system multiple times in every training epoch -- each time with a source specific conditioning. In this way, for \(K\) sources, the effective number of training iterations per epoch for this system will be at least \(K\) times that of the number of training iterations for a single model from type (a) system. As in \cite{meseguer2019conditioned}, there could be even higher number of training iterations for an epoch if this system is also conditioned on more than one source at a time. A type (c) system estimates all the sources simultaneously without any conditioning using a single multi-output model. It neither requires training over a data sample multiple times in an epoch nor does it involve any additional trainable parameters than a single model from type (a) system. A multi-task model could potentially perform even better than the dedicated models by learning from the extra mutual information across the tasks and sharing inductive bias as pointed out by Caruana \cite{caruana1997multitask}.

In this work, we train a single multi-channel U-Net (M-U-Net), a type (c) system, for multi-instrument source separation using a weighted multi-task loss function. We investigate the source separation task in two settings: 1) singing voice separation (two sources), and 2) multi-instrument source separation (four sources). The number of final output channels of our M-U-Net corresponds to the total number of sources in the chosen setting. Each loss term in our multi-task loss function corresponds to the loss on the respective source estimates. We explore Dynamic Weighted Average (DWA) \cite{liu2019end} and Energy Based Weighting (EBW) strategies to determine the weights for our multi-task loss function. We compare the performance of our M-U-Net trained with multi-task loss to that of dedicated U-Nets and the C-U-Net. Then, we investigate the effect of training with the silent-source samples\footnote{Data samples containing at least one silent source.} on the performance. We also study the effect of the choice of loss term definition on the source separation performance.

Our main contributions are: 
\begin{itemize}
    \item to propose M-U-Net as a computationally cheaper alternative (in terms of the number of training iterations and trainable parameters) for multi-instrument source separation to C-U-Net and the dedicated U-Nets.
    \item to propose a novel weighting strategy, EBW, for the multi-task loss function, based on the energy distribution in the ground truth sources.
    \item to show that training a model by discarding the data samples containing silent sources could reduce the overall number of training iterations and yet perform as good as the model trained with all the training data samples.
    \item to emphasize the importance of choosing appropriate signal representation for computing the loss term.
    
\end{itemize}

The rest of the paper is organized as follows. We review the related works in the context of our work in Section II. In Section III, we describe our source separation methodology. In Section IV, we explain our experimental setup and the experiments in detail along with the ablation studies. The final section is reserved for conclusion. The source code, along with the pre-trained model weights, audio examples and more elaborate tabulation of results are made available on \url{https://vskadandale.github.io/multi-channel-unet}

\section{Related Work}
In this section, we mainly focus on the U-Net based source separation methods. For the sake of clear comparison, we restrict the comparison of performance of our method with these U-Net based methods only. The objective of our work is not to achieve the best performance in source separation among all other approaches, but to highlight how a multi-task model could achieve performance comparable to that of a group of isolated single-task models at a much lesser training cost. Hence, we choose to work with a simple U-Net based model operating on spectrogram input. In our context, multi-task refers to group of parallellized single tasks where each task is to estimate a specific source, as in \cite{meseguer2019conditioned}.

Jansson et al. \cite{jansson2017singing} proposed using a pair of independently trained U-Nets (type (a) system in Fig.~\ref{fig1}) for the singing voice separation. Meseguer-Brocal and Peeters \cite{meseguer2019conditioned} pointed out that the implementations of such source-specific models get computationally expensive when there is a larger number of sources to be estimated. They proposed Conditioned U-Net (C-U-Net) as a cheaper alternative to the system of dedicated U-Nets, achieving comparable performance to that of the latter despite being a single model. The C-U-Net introduces control parameters through Feature-wise Linear Modulation (FiLM) layers in the encoder part of U-Net which adapts the model to estimate the source of the desired choice. C-U-Net corresponds to the type (b) system in Fig.~\ref{fig1}. As discussed earlier, the number of training iterations per epoch is much higher for a C-U-Net than a dedicated U-Net since every sample needs to be passed at least as many times as the number of sources to be estimated. Also, the inclusion of control parameters further adds to the training cost of C-U-Net. For these reasons, we investigate the possibility of using a single multi-channel U-Net (M-U-Net) (a type (c) system from Fig.~\ref{fig1}) which is computationally cheaper than the system of dedicated U-Nets and C-U-Net. Oh et al. \cite{oh2018spectrogram} propose a multi-channel U-Net (another type (c) system) for music source separation by adjusting the number of output channels to match the number of sources to be estimated. They also propose a weighting scheme for the multi-task loss function to balance the effect of unequal volume levels of different sources. Their network estimates the magnitude spectrograms directly as the output. In our work, we explore several weighting schemes for optimizing the multi-task loss function and compare the performance of our approach with that of dedicated U-Nets, C-U-Net and Oh et al. For the sake of fair comparison, we adapt implementation of Oh et al. and C-U-Net to our experimental setup.

\section{Proposed Method}
We train a Multi-channel U-Net (M-U-Net) that generates multiple outputs, one per source in the mixture. Having multiple outputs gives rise to multiple task-specific loss terms and hence the following multi-task loss function:
\begin{equation}
    \mathcal{L}=\sum_{i=1}^K w_i  L_i, \label{multitaskloss}
\end{equation}
where \(L_i\) is the loss term corresponding to the \(i\)-th source, \(w_i\) is its corresponding weight, \(K\) is the number of sources and \(\mathcal{L}\) is the overall scalar-valued loss. The input to our M-U-Net is the log-magnitude spectrogram of an audio mixture data sample. We train the U-Net to generate soft masks \(\hat{M}_i\) as the outputs. 

We explore two different definitions for the individual loss terms \(L_i\):
{\setlength{\parindent}{0cm}
\paragraph{\textit{Direct Loss}}
In this case, we first determine the Ideal Amplitude Masks (IAM \cite{wang2014training}) \(M_i\) for the ground truth source magnitude spectrograms \(S_i\) for each time-frequency bin \((n,m)\) as:
\begin{equation}
    M_i(n,m) = \min\left\{\frac{S_i (n,m)}{S_{mix} (n,m)},10\right\}, \label{iam}
\end{equation}
where \(S_{mix} \in \mathbb{R}_+^{F\times T}\) is the mixture magnitude spectrogram. We clip the values exceeding 10 for numerical stability in training. We then find the mean absolute value error (L1 loss) directly between the original source IAM masks \(M_i\) and their respective estimated masks \(\hat{M_i}\):
}
\begin{equation}
L_i = \sum_{n=1}^T \sum_{m=1}^F \lvert{M_i(n,m) - \hat{M_i}(n,m)}\rvert \label{directloss}
\end{equation}
{\setlength{\parindent}{0cm}
\paragraph{\textit{Indirect Loss}}
In this case, we find the mean absolute error between the original source spectrograms and the estimated spectrograms as shown in \eqref{indirectloss}. It is `indirect' in the sense that the U-Net outputs the masks but the loss term is defined on the spectrogram representations rather than the masks. This kind of loss term definition has been used in source separation works like \cite{zhao2018sound,meseguer2019conditioned}. Michelsanti et al. \cite{michelsanti2019training} showed that such an indirect loss performs better than the direct loss term for the speech enhancement task.
}
\begin{equation}
L_i = \sum_{n=1}^T \sum_{m=1}^F \lvert{S_i(n,m) - \hat{M_i}(n,m)\, S_{mix}(n,m)}\rvert \label{indirectloss}
\end{equation}


\subsection{Loss Weighting Strategies}\label{AA}
Now, we shift the focus on determining the weights \(w_i\) for each loss term \(L_i\) in \eqref{multitaskloss}. The ranges of loss values vary from one task to another. This results in competing tasks which could eventually make the training imbalanced. Training a multi-task model with imbalanced loss contributions might eventually bias the model in favor of the task with the highest individual loss, undermining the other tasks. Since all the tasks are of equal importance to us and the ranges of their individual loss terms differ, we cannot treat the loss terms equally. We need to assign weights to these individual loss terms indicative of their relative importance with respect to each other. Finding the right set of weights helps to counter the imbalance caused by the competing tasks during training and helps the multi-task system learn better. For determining the weights of losses in our multi-task loss function, we explore mainly the Dynamic Weight Average (DWA) and the Energy Based Weighting (EWB) strategies in this paper:

\subsubsection{Dynamic Weight Average (DWA)}
Liu et al. \cite{liu2019end} proposed the Dynamic Weight Average method for continuously adapting the weights of losses in a multi-task loss function during training. In this method, the weights are distributed such that a loss term decreasing at a higher rate is assigned a lower weight than the loss which does not decrease much. In this way, the model learns to focus more on difficult tasks rather than selectively learning easier tasks. The weight \(w_i\) for the $i$-th task  is determined as:
\begin{equation}
w_i(t) := \frac{K \exp{(\gamma_i(t-1)/T})}{\sum_j \exp{(\gamma_j(t-1)/T})}, \gamma_i(t-1)= \frac{L_i(t-1)}{L_i(t-2)}, \label{dwaeq}
\end{equation}
where $\gamma_i$ indicates the relative descending rate of the loss term \(L_i\), \(t\) is the iteration index, and \(T\) corresponds to the temperature which controls the softness of the task weighting. More the value of \(T\), more even the distribution of weights across all the tasks. 

In our work, we use DWA with \(T = 2\), the loss term \(L_i (t)\) being the average loss across the iterations in an epoch for the $i$-th task. Like in \cite{liu2019end}, we also set \(\gamma_i(t)=1\) for \(t=1,2\) to avoid improper initialization.

\subsubsection{Energy Based Weighting (EBW)}
\begin{figure}[ht]
\centering
\includegraphics[height=8cm,scale=0.5]{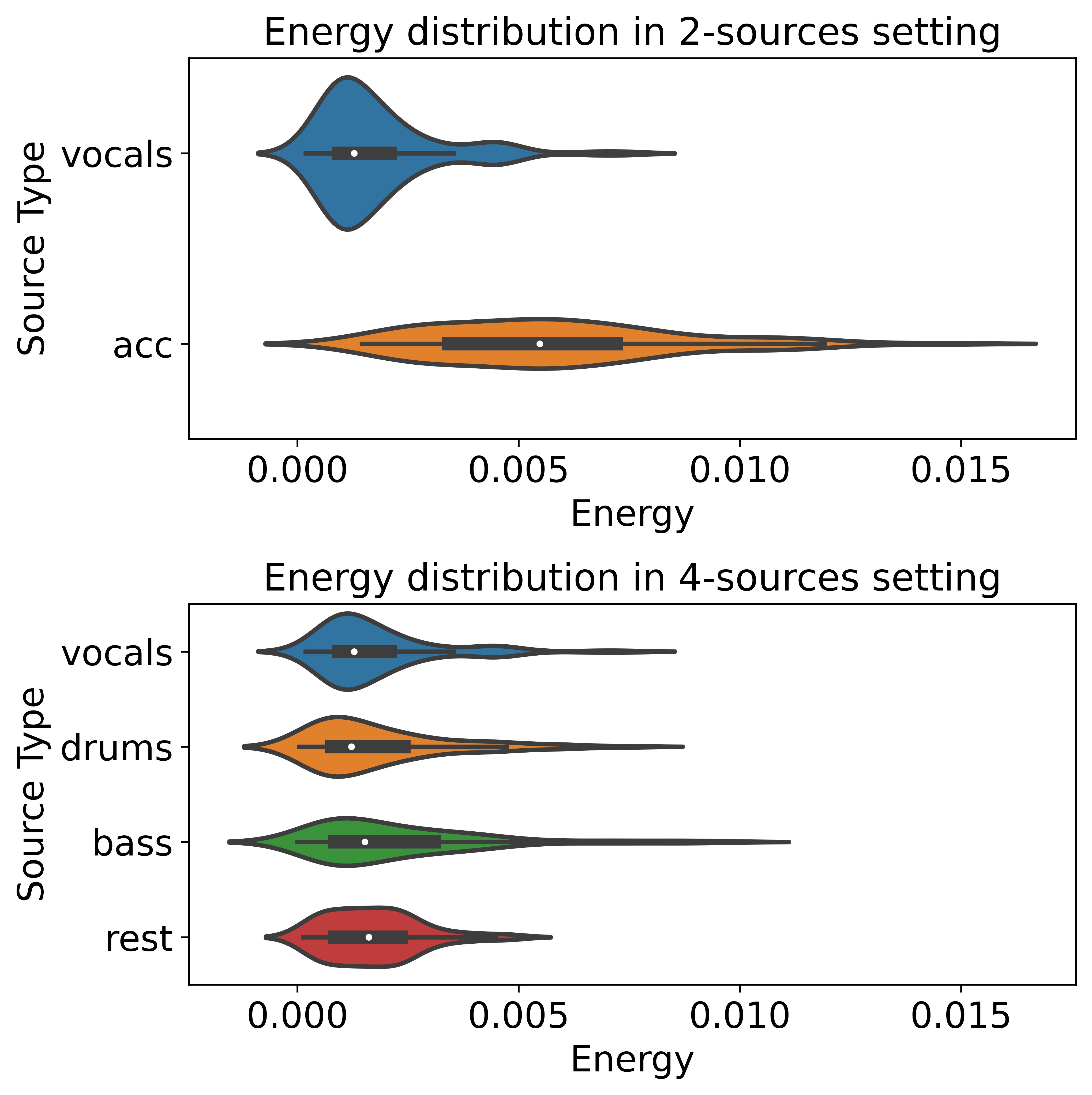}
\caption{Energy distribution across the sources.}
\label{energyprf}
\end{figure}

We determine the energy of a target source by summing the square value of each time-frequency bin in a magnitude spectrogram of a target source for a sample, normalize it by dividing by the number of time-frequency bins and then averaging across all the samples for the specific source. We notice that the energy distribution across the target sources is non-uniform (see Fig.~\ref{energyprf}). In the singing voice separation setting, the average energy of accompaniment is more than that of the vocals. In the multi-instrument source separation setting, the bass has relatively higher average energy than the other sources. We hypothesize that the uneven energy distribution could be a reason why the multi-task model preferentially learns certain tasks more than the others. When we trained our multi-task model with unit weighted loss, the estimates of sources with higher average energy were better than that of lower energy sources. Hence, we propose a weighting strategy based on energy distribution in the target sources such that the model does not become biased to the sources with higher energy.

We explore the following energy-based weighting settings in this work:
\paragraph{EBW\_P1} In this setting, we use the average energy content in \(i\)-th source, \(\mathrm{E}_i\), across all the samples (see \eqref{ebwp1}). Note that the weights are constant throughout the training for this setting.
\begin{equation}
w_i = \underset {j \in \{1, ..., K\}}{\max} \mathrm{E}_j\,/\,\mathrm{E}_i \label{ebwp1}
\end{equation}
This way, $w_i \geq 1$ for all tasks, being $w_i=1$ for the task associated to the source with highest energy and $w_i>1$ for the rest, and, in particular, the lower the energy of a specific source the higher its corresponding weight $w_i$, thus keeping the balance among tasks.  

\paragraph{EBW\_InstP1} In this setting, we use the average energy content in \(i\)-th source, \(\mathrm{E}_i\), across all the samples in a batch at iteration \(t\) as shown in \eqref{ebwinstp1}. Note that the weights change during the training for this setting.
\begin{equation}
w_i(t) = \underset {j \in \{1, ..., K\}}{\max} \mathrm{E}_j(t)\,/\,\mathrm{E}_i(t) \label{ebwinstp1}
\end{equation}

\paragraph{EBW\_P2} This setting is very similar to that of EBW\_P1 except for the fact that there is a power of 2 while determining the weights thus strengthening the relative importance between the task related to the highest-energy source and the rest of sources:
\begin{equation}
w_i = \underset {j \in \{1, ..., K\}}{\max} \mathrm{E}_j^2\,/\,\mathrm{E}_i^2 \label{ebwp2}
\end{equation}
Note that these weights are constant throughout the training for this setting as in EBW\_P1.

\paragraph{Oh et al. \cite{oh2018spectrogram}} Finally, we also experiment with the weighting scheme proposed by Oh et al. \cite{oh2018spectrogram}. In this setting, the weights are determined by solving the following pair of equations:
\begin{equation}
w_1 \mathrm{E}_1 = w_2\mathrm{E}_2 = ... = w_i\mathrm{E}_i = ... = w_K\mathrm{E}_K\label{ohetal1}
\end{equation}
\begin{equation}
\sum_{i=1}^K w_i = 1 \label{ohetal2}
\end{equation}


\section{Experiments}
In this section, we explore the effect of weighting strategies discussed in the previous section in training a multi-task model for source separation and compare their performance to that of a system of dedicated models and a conditioned multi-task model. 
We also perform ablation studies concerning the effect of choice of loss term  and the effect of silent-source samples.

\subsection{Dataset}
We use the Musdb18 \cite{musdb18} dataset for this work. It contains 150 full-length stereo (two channels) audio tracks along with the isolated constituent sources. The ground truth sources are available in two settings: i) 2 sources (vocals and accompaniment), and ii) 4 sources (vocals, drums, bass and rest). The dataset comes with a pre-defined split of 100 tracks for training and 50 for testing. We convert them to mono (single channel), downsample the audio to 10880Hz (as in \cite{zhao2018sound}) and split them into 6s long chunks without any overlap.  We then apply the Short-time Fourier Transform (STFT) on these chunks using a `Hanning' window of size of 1022 and hop-size of 256. This results in spectrograms of size 512\(\times\)256. We resample these spectrograms to 256\(\times\)256. These preprocessing steps are similar to that of \cite{zhao2018sound}. We move 5\% of the spectrogram samples from the training set to form our validation set. From this new training set, we filter out the silent-source samples.
\subsection{Network Architecture}

All the models in this work are based on a basic U-Net \cite{ronneberger2015u} model comprising of filters of sizes \{32, 64, 128, 256, 512, 1024, 2048\}. We have 6 down-convolution blocks, a transition block and 6 upconvolution blocks along with the skip connections in-between them. Throughout the experiments, we train the models using the Stochastic Gradient Descent (SGD) optimizer with a learning rate to 0.01 (unless otherwise mentioned) and a dropout of 0.1. The input to all our models is the log-magnitude spectrogram of audio mixture data sample of dimensions 256\(\times\)256.

In case of the dedicated U-Nets, we use a U-Net with single channel output since it estimates only a single source at a time. For the C-U-Net model, we adapt the implementation of C-U-Net provided by \cite{meseguer2019conditioned} to make it consistent with our U-Net architecture for a fairer comparison. In C-U-Net too, there is a single channel output as it estimates only one source at a time. With regard to Oh et al., we only test the performance of their weighting scheme in our experiment setup along with the other weighting schemes that we propose in this paper. For the sake of a fair comparison, we adapt the Oh et al. method to estimate masks instead of magnitude spectograms as in their original work. In our M-U-Net, the number of output channels corresponds to the total number of sources to be estimated, \(K\). The computational cost for each of these models is reported in  Table \ref{numparams}. Note that our M-U-Net has the least number of trainable parameters as well as the training iterations as compared to the others. More the training iterations longer the training time. Also, note that the samples are processed faster in M-U-Net than in C-U-Net during the inference.

\begin{table}[tb]
\caption{Computational Cost}
\resizebox{\linewidth}{!}{%
\begin{tabular}{|l|c|c|c|c|c|c|}
\hline
\multicolumn{1}{|c|}{} & \multicolumn{2}{c|}{Dedicated U-Nets} & \multicolumn{2}{c|}{{\color[HTML]{000000} C-U-Net}} & \multicolumn{2}{c|}{{\color[HTML]{000000} M-U-Net}} \\ \cline{2-7} 
\multicolumn{1}{|c|}{\multirow{-2}{*}{Model}} & 2src & 4src & 2src & 4src & 2src & 4src \\ \hline
\# params (approx.) & 248M & 496M & 162M & 162M & \textbf{124M} & \textbf{124M} \\ \hline
\begin{tabular}[c]{@{}l@{}}\# training iterations \\ per epoch\end{tabular} & N each & N each & 2N & 4N & \textbf{N} & \textbf{N} \\ \hline
\begin{tabular}[c]{@{}l@{}}processing rate \\ at inference (in sps) $^{*}$ \end{tabular} & 75 each & 84 each & 41 & 23 & \textbf{78} & \textbf{86} \\ \hline
\end{tabular}%
}
sps = approx. samples/second, N = number of training samples, M = million \\
\(^*\)metrics estimated for NVIDIA GeForce GTX 1070 GPU, batch size of 8
\label{numparams}
\end{table}

\subsection{Evaluation Metrics}
We choose to evaluate the following metrics \cite{vincent2006performance} : Source-to-Distortion Ratio (SDR),  Source-to-Interference Ratio (SIR) and Source-to-Artifact Ratio (SAR) typically used for evaluating music source separation performance. We use the mir\_eval toolbox \cite{raffel2014mir_eval} to get these metrics. Note that all our models estimate the soft masks and we obtain the magnitude spectrogram estimates of each source by multiplying these masks with the magnitude spectrogram of the mixture. We combine the phase of the mixture spectrogram along with the magnitude spectrograms of the estimated sources and apply inverse-STFT transform to obtain the waveforms. The metrics are evaluated on the waveforms of the estimated sources with respect to the appropriately downsampled ground truth audio waveforms. Among these three metrics, SDR is more indicative of the source separation quality as a global performance measure \cite{vincent2006performance} and some works (e.g. \cite{stoller2018wave}) report only this metric. We have not published the SAR and SIR  performance metrics in this paper for the sake of ease of readability. A more detailed tabulation of the results along with these metrics, is made available on the project webpage.

\subsection{Multi-task Experiments}
We aim to show that our M-U-Net can perform as good as the dedicated U-Nets for both singing voice separation (2 sources) and multi-instrument source separation (4 sources) with fewer trainable parameters. We conduct experiments with the M-U-Net exploring the weighting strategies: \{DWA, EBW\_P1, EBW\_InstP1, EBW\_P2 and Oh et al. \cite{oh2018spectrogram}\} which have been discussed earlier. We compare the performance of these M-U-Nets with the dedicated U-Nets and the C-U-Net. To notice the effectiveness of the weighting strategies, we also train an M-U-Net with unit weights (UW) and compare the performance with the models trained with our weighting strategies. Throughout the experiments, unless otherwise mentioned, we use the indirect loss function definition \eqref{indirectloss}. Table \ref{2srcresults} and Table \ref{4srcresults}, respectively, report the results for singing voice separation and multi-instrument source separation.

\begin{table}[tb]
\caption{Results of Singing Voice Separation in SDR (median in parenthesis)}
\resizebox{0.99\linewidth}{!}{%
\LARGE
\begin{tabular}{|c|c|c|c|}
\cline{1-4}
Model &  Vocals  &  Accompaniment &  Overall \\ \hline
\begin{tabular}[c]{@{}c@{}}Dedicated\\ U-Nets (x2)\end{tabular} & {5.09 \(\pm\) 4.31 (5.61)} & {12.95 \(\pm\) 3.18 (12.53)} & {9.02 \(\pm\) 5.46 (9.64)} 
\\ \hline \hline \hline

 C-U-Net &  4.42 \(\pm\) 4.98 (5.17) &  12.21 \(\pm\) 2.58 (12.16) &  8.31 \(\pm\) 5.56 (9.26) 
\\ 
\hline \hline

 UW  & 5.06 \(\pm\) 4.93 (5.75) &  12.98 \(\pm\) 3.14 (12.48) & 9.02 \(\pm\) 5.72 (9.74) 
\\ \hline \hline

DWA &  5.20 \(\pm\) 4.50 (5.67) & 12.96 \(\pm\) 3.11 (12.44) & 9.08 \(\pm\) 5.48 (9.61) 
\\ \hline \hline

 EBW P1 & 5.12 \(\pm\) 4.78 \textbf{(5.89)} &  \textbf{13.06} \(\pm\) 2.91 \textbf{(12.88)} & 9.09 \(\pm\) 5.60 (9.77) 
\\ \hline \hline

EBW InstP1 & \textbf{5.28} \(\pm\) 4.60 (5.79) & 13.04 \(\pm\) 3.02 (12.69) &  \textbf{9.16} \(\pm\) 5.50 \textbf{(9.79)}
\\ \hline \hline

Oh et al. \cite{oh2018spectrogram} & 5.18 \(\pm\) 4.17 (5.67) & 13.00 \(\pm\) 3.03 (12.63) &  9.09 \(\pm\) 5.35 (9.78)
\\ \hline

EBW P2$^{*}$ & 5.07 \(\pm\) 4.56 (5.63) & 12.89 \(\pm\) 2.95 (12.39) & 8.98 \(\pm\) 5.48 (9.66)
\\ \hline

\multicolumn{4}{l}{$^{*}$ trained with learning rate 0.001 instead of 0.01}

\end{tabular}%
}
\label{2srcresults}
\end{table}

From these tables, we notice that the source separation performance gets worse as the sources increase from 2 to 4, across all methods. In general, we find the performance of M-U-Nets trained using our weighting strategies comparable to that of Oh et al. method, C-U-Net and the dedicated U-Nets. Especially for the 2 source setting, the energy based methods and DWA method not only perform better than the C-U-Net and the Dedicated U-Nets, but also outperform the naive unit weighting (UW) based model, indicating the usefulness of our weighting strategies. As seen in Fig.~\ref{energyprf}, the average energy value between the sources differs a lot more in the 2 source setting than in the 4 source setting. Hence, the EBW methods which incorporate the signal energy information to weight the loss terms, perform better than the energy agnostic DWA method in the 2 source setting rather than in the 4 source setting. Looking at the \textit{Overall} SDR metrics (last column in the tables), we notice that our M-U-Net performs better than the Dedicated U-Nets, C-U-Net and Oh et al. method in both 2-source and 4-source settings at much lesser training cost compared to C-U-Net and Dedicated U-Nets. 


\begin{table}[tb]
\begin{center}
\caption{Results of Multi-instrument Source Separation in SDR (median values)}
\resizebox{0.7\linewidth}{!}
{%
\LARGE
\begin{tabular}{|c|c|c|c|c|c|}
\cline{1-6}
Model & Vocals & Drums & Bass & Rest & Overall  \\ \hline
\begin{tabular}[c]{@{}c@{}}Dedicated  \\ U-Nets (x4)\end{tabular} & 
\textbf{5.77} & 4.60 & \textbf{3.19} & 2.23 & 3.61
\\ \hline \hline \hline

C-U-Net & 5.26 & 4.30 & 2.97 & 1.69 & 3.37 
\\ \hline \hline

UW & 5.46 & 4.72 & 2.81 & 2.49 & 3.58
\\ \hline \hline

DWA & 5.24 & \textbf{4.92} & 2.88 & 2.45 & 3.61  
\\ \hline \hline 

EBW P1  & 5.41 & 4.77 & 2.94 & \textbf{2.64} & 3.65 
\\ \hline \hline 

EBW InstP1 & 5.46 & 4.85 & 2.86 & 2.58 & 3.52
\\ \hline \hline 

Oh et al. \cite{oh2018spectrogram} & 5.29 & 4.86 & 2.85 & 2.55 & 3.60
\\  \hline \hline

EBW P2 & 5.44 & 4.89 & 2.99 & 2.58 & \textbf{3.66}
\\ \hline

\end{tabular}%
}
\label{4srcresults}
\end{center}
\end{table}

\subsection{Ablation Studies}
Now, we present some additional experiments to evaluate which loss definition, among direct loss and indirect loss, performs better. We also analyze the effect of including silent-source samples in the training set. For these additional experiments, we consider the setting EBW\_P1 on 4 sources as a reference. Table \ref{additional} reports the performance metrics pertaining to these experiments.

\begin{table}[t]
\caption{Results of Ablation Studies (median in parenthesis)}
\resizebox{\linewidth}{!}{%
\begin{tabular}{|c|c|c|c|}
\hline
\multirow{2}{*}{Model} & \multicolumn{3}{c|}{Overall Performance Metrics} \\ \cline{2-4} 
& SDR & SIR & SAR \\ \hline
EBW\_P1$^{*}$ & {\bf 3.46} \(\pm\) 4.15 {\bf (3.65)} & 7.97 \(\pm\) 5.04 (7.93) & {\bf 6.83} \(\pm\) 3.18 {\bf (6.77)} \\ \hline
\begin{tabular}[c]{@{}c@{}}EBW\_P1 with \\ Direct Loss \eqref{directloss} \end{tabular} & 3.31 \(\pm\) 3.96 (3.46) & 7.98 \(\pm\) 4.97 (8.18) & 6.64 \(\pm\) 3.12 (6.67) \\ \hline
\begin{tabular}[c]{@{}c@{}}EBW\_P1$^{*}$  \\ without filtering\end{tabular} & 3.44 \(\pm\) 4.28 (3.59) & {\bf 8.33} \(\pm\) 5.10 {\bf (8.24)} & 6.61 \(\pm\) 3.46 (6.72) \\ \hline
\multicolumn{4}{l}{$^{*}$ trained using Indirect Loss \eqref{indirectloss}}
\end{tabular}
}
\label{additional}
\end{table}

From Table \ref{additional}, we notice that the overall performance drops in both the ablation studies. Despite the slight improvement in the SIR metric on using direct loss \eqref{directloss}, based on the SDR and SAR metrics, we recommend using indirect loss \eqref{indirectloss} definition which is congruent with the findings for speech enhancement \cite{michelsanti2019training}. We also infer that training with silent-source samples does not contribute much to the overall performance and we recommend discarding them. 

\section{Conclusion}
We presented a multi-channel U-Net as a cheaper alternative (in terms of the number of training iterations and trainable parameters) to the Conditioned U-Net and the system of dedicated U-Nets for music source separation. It is also considerably faster than the Conditioned U-Net at inference. Such an approach could be potentially extended to models other than the U-Net and perhaps also for other kinds of tasks. We also presented a novel weighting strategy, EBW, for training the multi-task loss function based on the energy of the signal representations. We showed how the EBW method is effective when the average energy values across the sources to be estimated are very different. We believe there are other ways of distilling the energy information into the weighting strategy and leave it for future work. We also showed that discarding silent-source samples during training saves on the training cost without much compromise in performance. We also showed that M-U-Net performs better when trained with an indirect loss term rather than the direct loss on the masks.

\section*{Acknowledgment}
We thank Daniel Michelsanti (Aalborg University) and Olga Slizovskaia (Universitat Pompeu Fabra) for the insightful discussions related to source separation methods and practices.

\bibliographystyle{IEEEtran}
\bibliography{IEEEabrv,mybib}

\end{document}